\documentclass[twocolumn]{aastex61}
\usepackage[utf8]{inputenc}
\usepackage{txfonts}
\usepackage{natbib}
\usepackage{graphicx}
\usepackage{amssymb}
\usepackage{soul}

\def\orten{2007\,OR\ensuremath{_{10}}}
\def\ortenlong{(225088)~2007\,OR\ensuremath{_{10}}}
\def\geomalb{p$_{\mathrm{V}}$}
\def\absmag{H$_{\mathrm{V}}$}

\shorttitle{The satellite of \orten}
\shortauthors{Kiss et al.}

\begin{document}
\sloppy

\title{Discovery of a satellite of the large trans-Neptunian object \ortenlong}

\correspondingauthor{Csaba Kiss}
\email{kiss.csaba@csfk.mta.hu}

\author[0000-0002-8722-6875]{Csaba Kiss}
\affil{Konkoly Observatory, Research Centre for Astronomy and Earth Sciences, Hungarian Academy of Sciences, Konkoly Thege 15-17, H-1121~Budapest, Hungary}

\author[0000-0002-1326-1686]{G\'abor Marton}
\affiliation{Konkoly Observatory, Research Centre for Astronomy and Earth Sciences, Hungarian Academy of Sciences, Konkoly Thege 15-17, H-1121~Budapest, Hungary}

\author{Anik\'o Farkas-Tak\'acs}
\affiliation{Konkoly Observatory, Research Centre for Astronomy and Earth Sciences, Hungarian Academy of Sciences, Konkoly Thege 15-17, H-1121~Budapest, Hungary}

\author{John Stansberry}
\affiliation{Space Telescope Science Institute, 3700 San Martin Dr., Baltimore, MD 21218, USA}

\author{Thomas M\"uller}
\affiliation{Max-Planck-Institut für extraterrestrische Physik, Postfach 1312, Giessenbachstr., D-85741 Garching, Germany}

\author[0000-0001-8764-7832]{J\'ozsef Vink\'o}
\affiliation{Konkoly Observatory, Research Centre for Astronomy and Earth Sciences, Hungarian Academy of Sciences, Konkoly Thege 15-17, H-1121~Budapest, Hungary}
\affiliation{Department of Optics and Quantum Electronics, University of Szeged, D\'om t\'er 9, H-6720 Szeged, Hungary}

\author{Zolt\'an Balog}
\affiliation{Max-Planck-Institut f\"ur Astronomie, K\"onigstuhl 17, D-69117 Heidelberg, Germany}

\author{Jose-Luis~Ortiz}
\affiliation{Instituto de Astrof\'{i}sica de Andaluc\'{i}a - CSIC, Apt 3004, E-18080 Granada, Spain}

\author[0000-0001-5449-2467]{Andr\'as P\'al}
\affiliation{Konkoly Observatory, Research Centre for Astronomy and Earth Sciences, Hungarian Academy of Sciences, Konkoly Thege 15-17, H-1121~Budapest, Hungary}




\begin{abstract}

\orten{} is currently the third largest known dwarf planet in the
Transneptunian region, with an effective radiometric diameter of $\sim$1535\,km.
It has a slow rotation period of $\sim$45\,h that was suspected to be caused by tidal interactions with a satellite undetected at that time. Here we report on the discovery of a likely moon of \orten, identified on archival Hubble Space Telescope WFC3/UVIS system images. Although the satellite is detected at two epochs, this does not allow an unambiguous determination of the orbit and the orbital period. A feasible 1.5-5.8$\cdot$10$^{21}$\,kg estimate for the system mass leads to a likely 35 to 100\,d orbital period. The moon is about 4\fm2 fainter than \orten{} in HST images that corresponds to a diameter of 237\,km assuming equal albedos with the primary. 
Due to the relatively small size of the moon the previous size and albedo estimates for the primary remains unchanged. With this discovery all trans-Neptunian objects larger than 1000\,km are now known to harbour satellites, an important constraint for moon formation theories in the young Solar system. 

\end{abstract}

\keywords{      methods: observational --- 
                techniques: photometric --- 
                astrometry --- 
                minor planets, asteroids: general --- 
                Kuiper belt objects: individual (2007OR10)}


\section{Introduction} \label{sect:intro}

(225088) \orten{} (\orten{} hereafter for short) is a large (D\,$\approx$\,1500\,km) and distant (currently at r$_{\rm helio}$=87\,AU) trans-Neptunian object (TNO). In a recent study, \citet{Pal2016} analysed light curves of \orten{} obtained with the K2 mission of the Kepler Space Telescope. They found that \orten{} rotates very slowly relative to other trans-Neptunian objects, with a most likely period of P$_{rot}$\,=\,44.81$\pm$0.37\,h.
The canonical explanation of slow rotation for large bodies is tidal interaction with a fairly massive satellite. As discussed in \citet{Pal2016} the rotation period of \orten{} suggests that the suspected moon would be at an apparent separation of 0\farcs04--0\farcs08 assuming tidal locking and depending on their mass ratio. However, a smaller satellite at a larger separation could have slowed down the rotation of \orten{} to the observed value, but may not have been massive enough to force synchronous rotation. 

Assuming that the primary is the only notable body in the system the integrated thermal emission indicates that \orten{} has a diameter of 1535$^{+75}_{-225}$\,km, making it the third largest dwarf planet, after Pluto and Eris \citep{Pal2016}. With this diameter, \orten{} is larger than the officially recognised dwarf planets Makemake and Haumea. If a large satellite is present, the diameter of the primary could be correspondingly smaller. To date no satellite or binarity of \orten{} has been reported in the literature.

Motivated by these questions, we have checked \orten{} observations in the Hubble Space Telescope Archive and identified a likely satellite. In this letter we describe the putative moon's characteristics as derived from these observations.  
\begin{figure*}[ht!]
\begin{center}
\includegraphics[width=\textwidth]{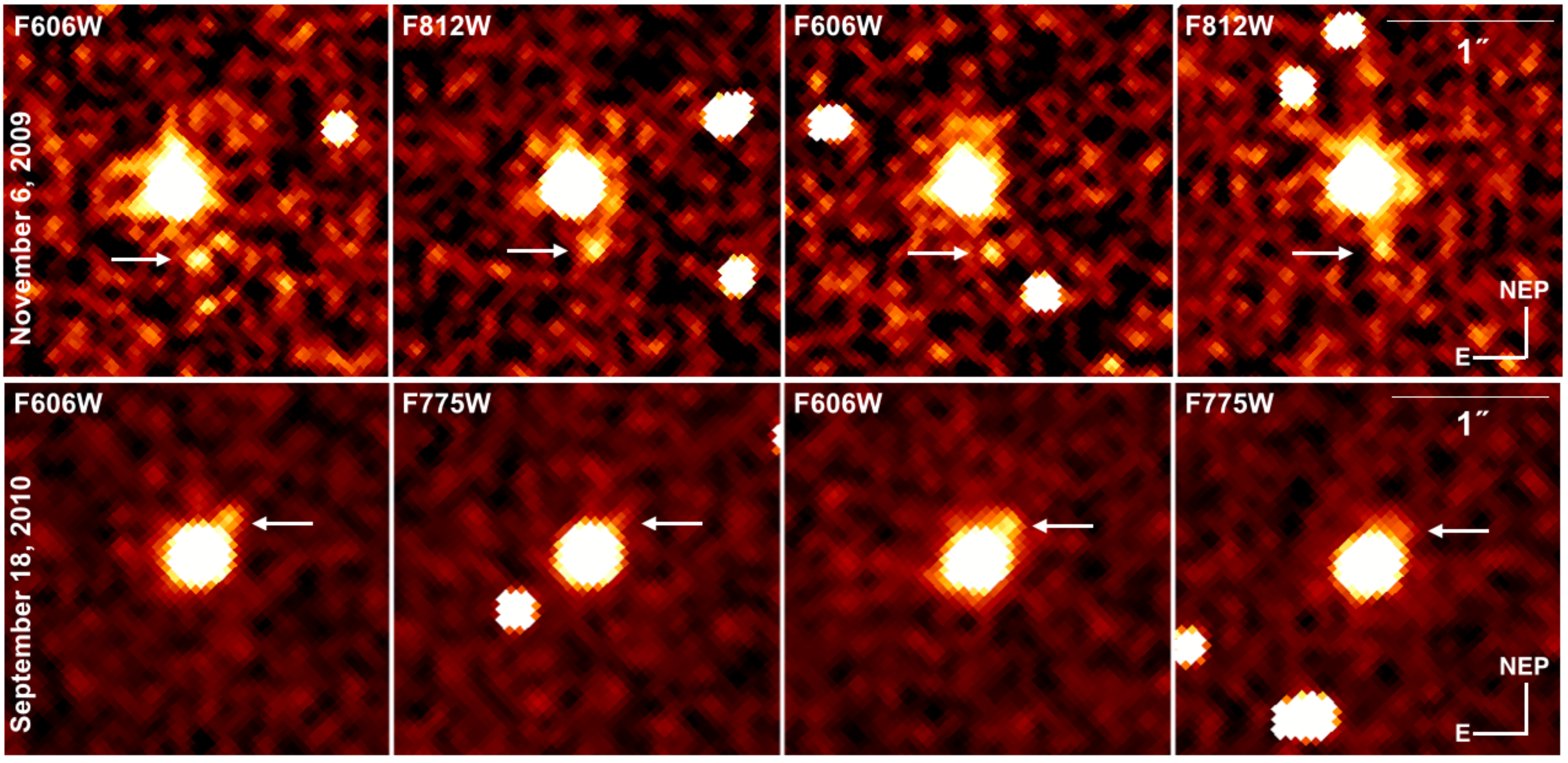} 
\end{center}
\caption{Hubble Space Telescope WFC3/UVIS images of \orten. Upper raw: November 6, 2009 measurements, F606W-F812W-F606W-F812W filter series; botom raw: 
September 2010 images, F606W-F775W-F606W-F775W filter series. The suspected satellite can be most readily identified on the F606W images and is marked by a white arrow on each image (North is up and East is left, in Ecliptic coordinates). 
\label{fig:images}}
\end{figure*}
\begin{figure*}[ht!]
\begin{center}
\includegraphics[width=\textwidth]{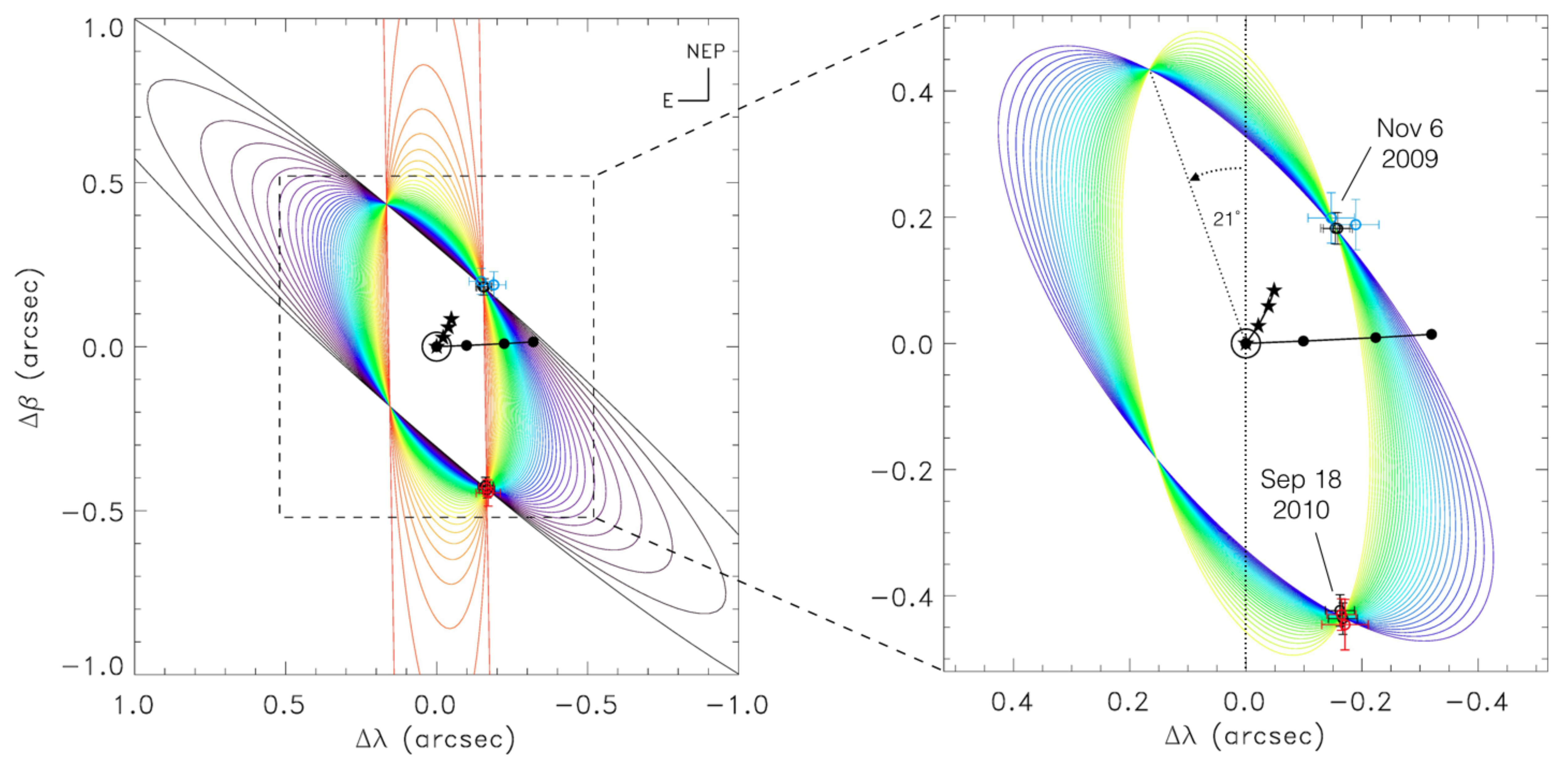}
\end{center}
\caption{Relative positions of two sources, one at each epoch, identified as the putative satellite, in ecliptic coordinates, with respect to \orten{}. The sources are marked by open circles with error bars, \orten{} is the large open circle in the center. The colours of the satellite positions correspond to different filters: F606 - black; F755W - blue; F812W - red. The sources co-move with \orten{} in the two series of images, i.e. their positions are the same in this figure, within the uncertainties. Curves connecting black dots and stars starting from the center indicate the movement of \orten{} with respect to the background between the exposures during the November 6, 2009 (stars) and September 18, 2010 images (filled circles). We plot the ellipses best fit to the observed positions of the source, assuming that \orten{} is in the centre (projection of circular orbits). 
The right panel is a magnified view of the innermost $\sim$1\arcsec$\times$1\arcsec region, showing only those ellipses that fit into the window \label{fig:orbits}}
\end{figure*}

\section{Observations and data analysis} \label{sect:obsanddata}

\subsection{Archival Hubble Space Telescope Observations} \label{sect:obs}

\orten{} was observed with the Hubble Space Telescope at two epochs, on November 6, 2009 (proposal ID: 11644, PI: M.~Brown) and on September 18, 2010 (proposal ID: 12234, PI: W.~Fraser). Both proposals used similar strategies, observing the target with a set of visual range and near-infrared filters of the WFC3/UVIS and IR cameras. Due to the better spatial resolution visual range observations are preferred in identifying unknown satellites, and we used the WFC3/UVIS observations to look for potential moons of \orten{} in these series of measurements. 

At the first epoch (November 6, 2009, 17:08:36 start time) \orten{} was observed with the WFC3/UVIS camera system using the 512-pixel sub-array mode with the UVIS1-C512A-SUB aperture, in a series of four measurements with the F606W-F814W-F606W-F814W filters. Each measurement lasted for 129\,s. A similar strategy was followed at the second epoch (September 18, 2010, 15:54:12 start time), now taking four measurements with the UVIS2-C512C-SUB aperture and using the F606W-F775W-F606W-F775W filter combination. The F606W measurements lasted for 128\,s, while the length of the F775W measurements were 114\,s (see also Table~\ref{table:summary}). 

There is a faint source in the vicitnity of \orten{} that appears in both epochs and in all images, and at the same location with respect to \orten{} at each epoch (see Table~\ref{table:summary} and Fig.~\ref{fig:images}). 

We used the drizzle images and routines built on the DAOPHOT-based APER function in IDL\footnote{Interactive Data Language, Harris Geospatial Solutions} to obtain aperture photometry and astrometry of the photocenters of both \orten{} and the suspected satellite. On the November 6, 2009 images aperture photometry could be performed for both targets separately, in both bands (F606W and F814W). In the case of the September 18, 2010 images, however, the satellite was too close to \orten{} and reliable photometry of the moon could only be performed after the subtraction of \orten's point spread function (PSF). This was modeled using the TinyTim \citep{krist2010} software, using specific setups of date, camera system, target's pixel position, focal length, and spectral energy distribution of the target (black body of 5800\,K). The TinyTim-created drizzle model images were adequate to subtract the contribution of \orten{} from the original drizzle images. The best-fit parameters of the model PSF were determined using Levenberg-Marquardt nonlinear least-square fitting. The extracted relative positions of the satellite are listed in Table~\ref{table:summary}. 

At the first observational epoch, \orten{} moved with an average
apparent velocity of \mbox{$\mathrm{v_\lambda}$\,=\,--0.33\,\arcsec\,h$^{-1}$} and \mbox{$\mathrm{v_\beta}$\,=\,--0.47\,\arcsec\,h$^{-1}$} in Ecliptic
longitude and latitude. The total motion observed in the sequence
of exposures was 0\farcs10 (2.5 pixels). At the second epoch, the
apparent velocities were \mbox{$\mathrm{v_\lambda}$\,=\,--1.86\,\arcsec\,h$^{-1}$} and \mbox{$\mathrm{v_\beta}$\,=\,0.01\,\arcsec\,h$^{-1}$}, and
the total observed motion was $\sim$0\farcs33 (8 pixels). Within each epoch,
the position of the secondary source relative to \orten{} was
constant to within the measurement errors of our astrometry (see
Fig.~\ref{fig:images}). Since those astrometric errors ($\sim$0\farcs04) are much smaller
than the observed motion of \orten, we confirm that the secondary
source was comoving at both epochs.

\begin{table*}
\caption{Summary table of the derived satellite characteristics as observed on the dates (start times) and with the filters given below. The table also lists the integration times (t$_{int}$), the brightness difference with respect to \orten{} ($\Delta$m), the offset in Ecliptic coordinates relative to \orten{} ($\Delta\lambda$,$\Delta\beta$), the heliocentric ($r_{\rm h}$) and geocentric distances ($\Delta$) and the phase angle ($\alpha$) at the time of the observations.   }
\begin{center}
\begin{tabular}{ccccccccc}
\hline
Epoch & Filter & t$_{int}$ & $\Delta$m & $\Delta\lambda$& $\Delta\beta$ & r$_{\rm h}$ & $\Delta$ & $\alpha$ \\
(JD)  &        & (s)       & (mag)     &    (\arcsec)   & (\arcsec)     & (AU) & (AU) & (deg)\\ 
\hline
2455142.2136 & F606W & 128 & 4.25$\pm$0.28 & -0.166$\pm$0.025 & -0.436$\pm$0.025 & 85.960 & 85.683 & 0.63\\
2455142.2159 & F814W & 128 & 4.30$\pm$0.30 & -0.164$\pm$0.025 & -0.429$\pm$0.025 \\
2455142.2188 & F606W & 128 & 4.61$\pm$0.29 & -0.162$\pm$0.025 & -0.423$\pm$0.025 \\
2455142.2211 & F814W & 128 & 4.43$\pm$0.38 & -0.170$\pm$0.040 & -0.445$\pm$0.040 \\
\hline
2455458.1619 & F606W & 129 & 4.13$\pm$0.18 & -0.154$\pm$0.025 & 0.183$\pm$0.025 & 86.175 & 85.263 & 0.27 \\
2455458.1642 & F775W & 114 & 4.64$\pm$0.30 & -0.189$\pm$0.040 & 0.188$\pm$0.040 \\
2455458.1669 & F606W & 129 & 4.17$\pm$0.19 & -0.158$\pm$0.025 & 0.182$\pm$0.025 \\
2455458.1692 & F775W & 114 & 4.31$\pm$0.23 & -0.147$\pm$0.040 & 0.199$\pm$0.040 \\ \hline
\end{tabular}
\end{center}
\label{table:summary}
\end{table*}

We also determined the brightness difference between \orten{} and its moon for each measurement (see Table~\ref{table:summary}). 
As in the case of relative astrometry, proper photometry was only possible after subtracting the PSF of the primary in the second epoch images. 

The uncertainties in the relative brightness determination reflect the low signal-to-noise ratio of the satellite detection, especially at the first epoch, when we detected it at the 3-4\,$\sigma$ significance level. There is a notable change in the brightness ($\sim$0\fm3) of the satellite relative to \orten between the two epochs. As the light curve of \orten{} is shallow \citep{Pal2016}, only a maximum of $\sim$0\fm09 difference can be attributed to the rotation of the primary. However, shape and/or albedo variegations on the surface of the satellite can easily account for the remaining flux difference. 
The mean brightness differences 
are found to be $\Delta{m}$(F606W)\,=\,4\fm23$\pm$0\fm24, $\Delta{m}$(F775W)\,=\,4\fm43$\pm$0\fm30 and $\Delta{m}$(F814W)\,=\,4\fm35$\pm$0\fm25. As  these are nearly equal in all bands, \orten{} and its satellite have very similar colors from the V to the I bands, roughly covered by the three HST/WFC3 filters used. We find it very unlikely that two independent, co-moving sources with similar brightness and both having the same color as \orten{} would be found in the vicinity of \orten{} at two epochs. Therefore we hypothesize that the two sources we found at the two epochs are two appearances of the same satellite.  

With these colours both \orten{} and the satellite are among the reddest objects known in the trans-Neptunian region. 

In general, red TNOs are seen to have higher albedos than gray objects \citep[see][]{Lacerda2014}. Since both \orten{} and the satellite are extremely red, our data suggest that they are likely to have similar albedos, and that the albedo of
\orten (\geomalb\,=\,0.089) probably applies to the satellite as well.


For the \orten{} system we adopt the absolute magnitudes and colours found in \citet{Hermann}, i.e. H$_{\rm V}$\,=\,2\fm34, H$_{\rm R}$\,=\,1\fm49, B-V\,=\,1\fm38, V-R\,=\,0\fm86, R-I\,=\,0\fm79. Consideration of the contribution of the satellite to the total brightness of the system increases the absolute brightness magntiude of \orten{} by $\sim$0\fm03, while the colors are nearly unchanged. This results in \mbox{H$_{\rm V}$\,=\,6\fm57$\sim$0\fm26} for the satellite. We use this value in the size and thermal emission calculations below. 

\subsection{Possible orbits of the satellite}

The two set of observations allowed us to set some constraints on the orbit of the satellite around \orten. We assume that the orbit of the satellite is circular as circularisation times are typically significantly shorter than the age of these systems \citep{Noll2008}. Then, the apparent ellipse of the orbit is a projection of the circular orbit, with \orten{} in the center in a co-moving frame. The two orbital positions defined by the two set of observations do not determine the orbit unambiguously, but allow a family of ellipses to be fitted, as presented in Fig.~\ref{fig:orbits}. In our case the possible position angles of the ellipses range from 1\degr{} to 51\degr{} 
(from North to East in Ecliptic coordinates). The semi-major and semi-minor axes of the smallest ellipse are 0\farcs46 and 0\farcs22 (29300 and 13600\,km) with 21\degr{} position angle. For smaller and larger values within the 1\degr{} to 51\degr{} range the semi-major axes increase quickly and get infinitely large at the limiting position angles. 

A reliable estimate for the mass of \orten{} can be obtained using the size limits of the thermophysical model calculations \citep{Pal2016}, D$_{\rm eff}$\,=\,1310--1610\,km. As \orten{} is a fairly large object, internal porosity is likely negligible and a lower limit for the density can be set to 1.2\,g\,cm$^{-3}$, a typical value for medium size TNOs \citep{Brown2013,Barr2016,Kovalenko}. For an upper limit we use the densities of the largest dwarf planets Pluto and Eris, and adopt 2.5\,g\,cm$^{-3}$. With these assumptions the mass of \orten{} would be 1.5--5.8$\cdot$10$^{21}$\,kg. Then, with the smallest possible semi-major axis the orbital periods would be 18\fd5--36\fd4, depending on the system mass assumed. 

\begin{figure}[ht!]
\includegraphics[width=8.7cm]{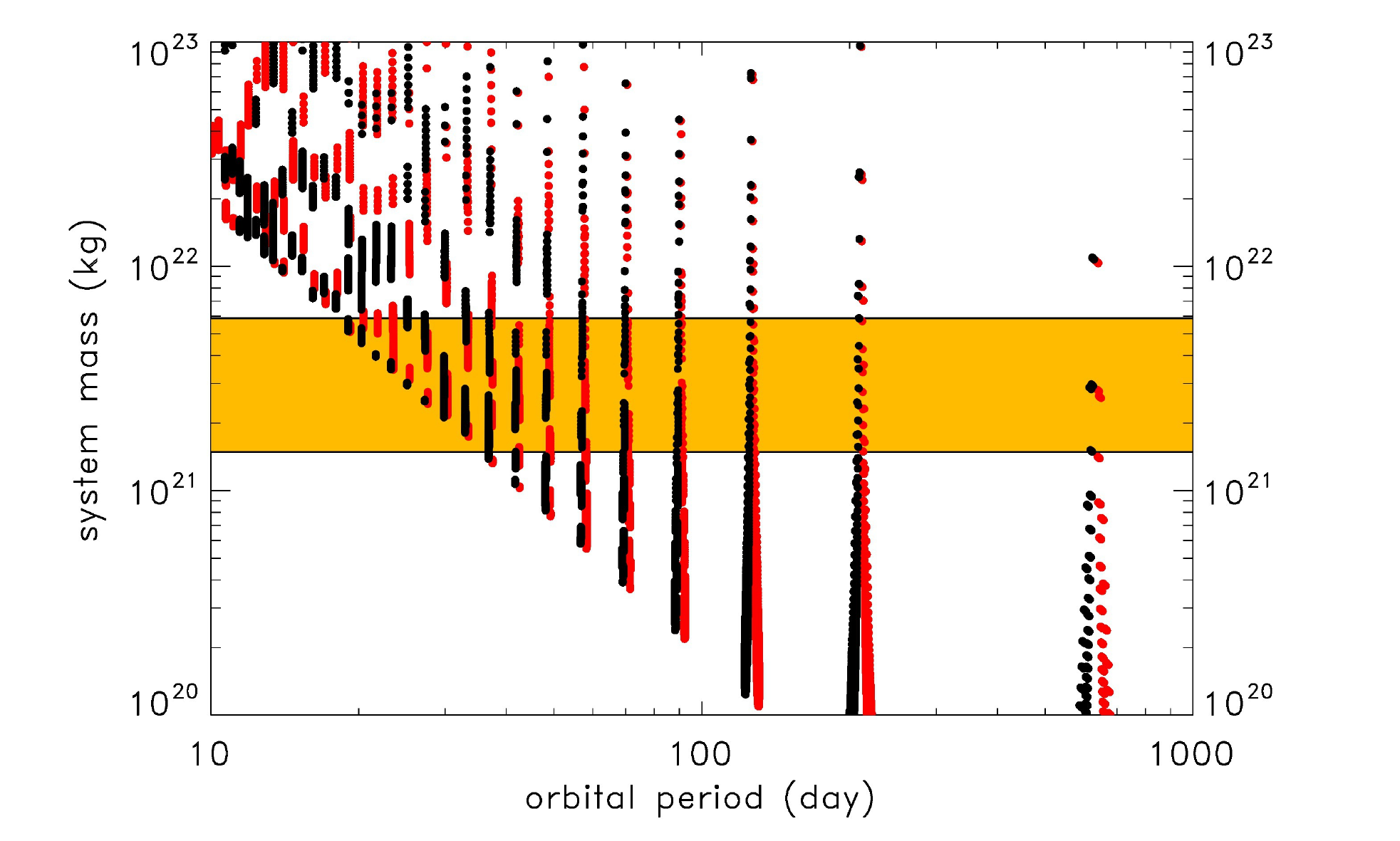}
\caption{Possible system mass values based on the fitted orbits and the observed orbital phase differences. The orange region shows the mass range based on an assumed range of 
density and size of \orten. Possible prograde and retrograde solutions are marked by black and red dots.  
\label{fig:thermal}}
\label{fig:mass}
\end{figure}

The two observed positions also define the orbital phases for a specific orbit (ellipse), and the phase difference can be used to find those orbital periods that are compatible with the observed positions, considering the time spent between the two set of observations (315\fd95). The semi-major axis and the orbital period also defines the system mass according to Kepler's third law. We applied this scheme to all ellipses fitted to the two satellite positions, determined the compatible orbital periods and calculated the related systems mass values. 
The results are presented in Fig.~\ref{fig:mass}. The shortest orbital periods compatible with the phase differences for any of the fitted ellipses are 19\fd05 for prograde (black dots) and 19\fd23 for retrograde (red dots) sense of revolution. Shorter orbital periods would require a mass too high for our upper limits (upper-left corner in Fig.~\ref{fig:mass}). Although only some well-defined orbital periods are allowed there, there are several of these possible orbital period groups in the 20 to 100\,day range. This means that neither the orbital period nor the system mass can be constrained further by the two existing HST observations. Although they cannot be fully excluded, orbital periods longer than $\sim$100\,d become increasingly unlikely as the satellite would spend most of the time at large apparent distances.
We have found three groups of possible periods at $\sim$126, $\sim$210 and $\sim$630\,d, but no additional orbital periods were identified for $>$1000\,d. 

The expected orbital period of a satellite can be estimated using the formalism in \citet{Murray}, assuming tidal dissipation and requiring that the current semi-major axis is significantly different from the initial one. 
In this case the orbital period is 
P\,$\propto$\,$k^{3/13}$$Q^{-3/13}$$q^{-3/13}$$m_p^{-5/13}$, where $k$ is the tidal Love number, $Q$ is the quality factor of the primary, $q$ is the ratio of the primary to the satellite mass and $m_p$ is the mass of the primary. With some reasonable assumptions for these parameters \citep[see also][]{Brown+Schaller,Brown2007}, and assuming an evolution of 4.5\,Gyr we can estimate the possible orbital periods. In the equal albedo and equal density case the mass ratio is $q$\,$\approx$\,350, and the high and low mass limits for the primary gives orbital periods between 45 and 76\,days. Orbital periods around 35\,days require a significant ($>$2.5$\times$) internal density difference between the primary and the moon. This is, however, reasonable concerning the known higher densities of the largest and the mid-sized trans-Neptunian objects \citep[see e.g.][]{Brown2013,Kovalenko}. In the case of a low albedo moon (\geomalb\,$\approx$\,5\%) and a low mass primary the orbital period would be P\,$\approx$\,100\,d. These calculations show that the preferred orbital periods are in the range of 35 to 100\,d, and that the orbits with the smallest semi-major axes and shortest periods (P\,$\approx$\,20\,d) may not be the most likely ones. 
 
\section{Thermal emission of the system} \label{sect:thermal}

\begin{figure}[ht!]
\includegraphics[width=8.7cm]{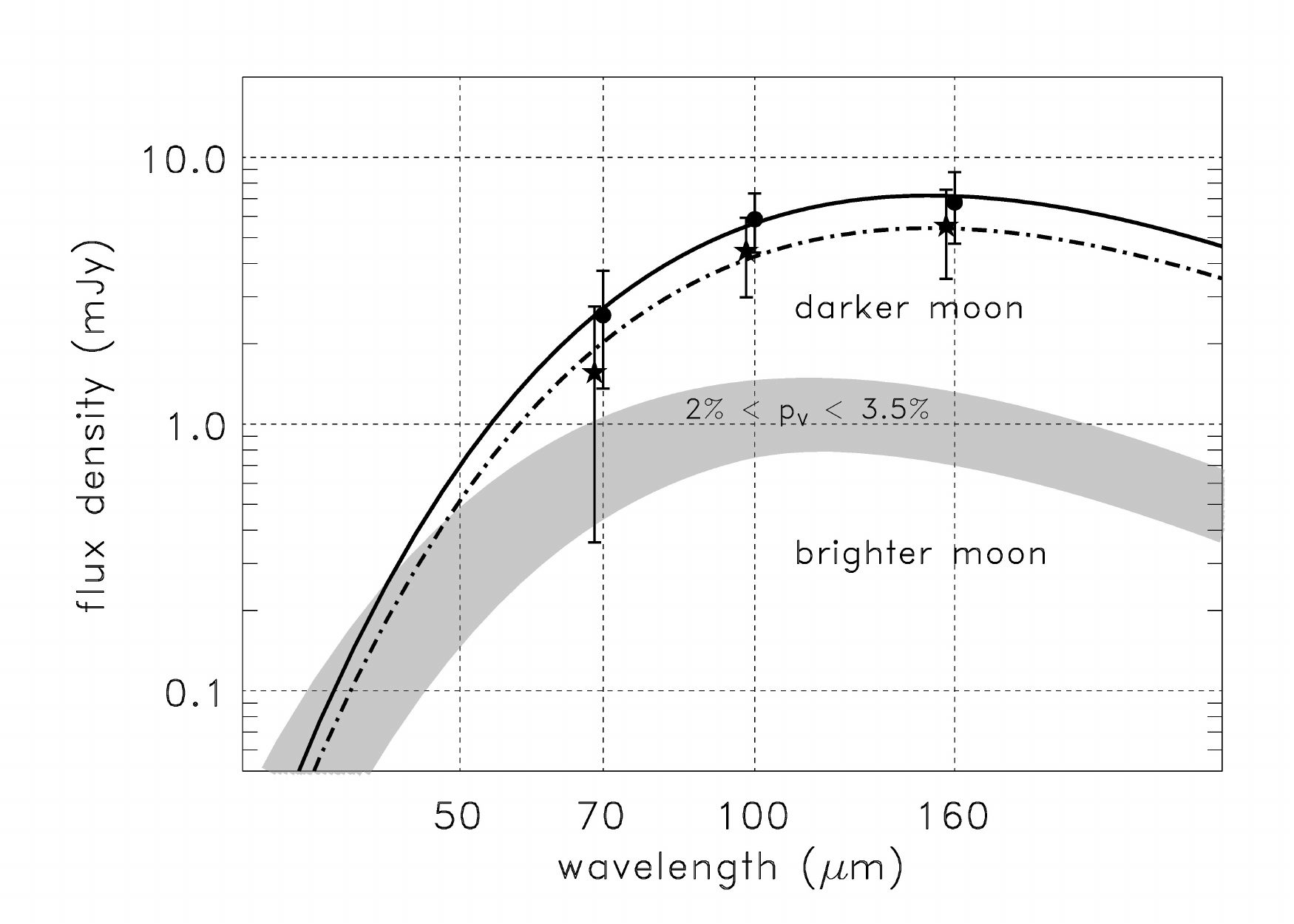}
\caption{Thermal emission components in the \orten{} system. The black dots with error bars represent the measured Herschel/PACS fluxes. The thick black curve
is the best-fit NEATM model with D\,=\,1535km and $\eta$\,=\,1.8 for the primary \citep{Pal2016}. The gray area represent NEATM thermal emission models of the putative satellite assuming geometric albedos in the range of 2\% to 3.5\% (size of ~350km), with very low ($\eta$\,$<$\,0.8) beaming parameter values. Stars with error bars and the related dash-dotted curve best-fit NEATM model represent the corrected thermal emission of \orten{} assuming an extremely dark moon (data points are slightly shifted in wavelength for clarity, see the text for details). }
    \label{fig:thermal}
\end{figure}


In the case of Makemake, a dwarf planet of similar size, the satellite may have a significant contribution to the thermal emission of the system due to the possibly large albedo difference 
\citep{Lim,Parker2016}. 
In the case of \orten{}, however, the primary is rather dark: \geomalb\,=\,0.089$_{-0.009}^{0.031}$ \citep{Pal2016}. We calculated the possible contribution of the satellite to the thermal emission using the Near-Earth Asteroid Thermal Model model (\citep[NEATM ][]{Harris}) assuming geometric albedos in the range of 2\% to 9\% for the satellite. We used the absolute magnitude of \absmag\,=\,6\fm57$\pm$0\fm26 determined above and applied the formula by \citet{Brucker} to obtain the phase integral and the Bond albedo. The upper limit of \geomalb\,=\,9\% we considered is the geometric albedo of the primary: in the case of higher albedos the contribution of the satellite would be negligible due to the large primary to satellite area ratio ($>$\,40). We allowed the beaming parameter $\eta$ to vary in the range of 0.6--2.5 
\citep[see e.g. ][]{Lellouch2013}. The far-infrared flux densities of the system, as observed with Herschel/PACS at 70, 100 and 160\,$\mu$m, are taken from \citet{Pal2016}.

As presented in Fig.~\ref{fig:thermal}, only extremely dark (\geomalb\,=\,2--3.5\%) and rough ($\eta$\,$<$\,0.8) surfaces provide a noticeable contribution to the total thermal emission. While such surfaces exist among Solar system bodies \citep[e.g.][]{Pal2015} the geometric albedos in the trans-Neptunian region 
are typically higher than this. The dark-neutral population of objects \citep{Lacerda2014} have typical geometric albedos of \geomalb\,$\approx$\,5\%, but practically no objects show \geomalb\,$<$\,4\%. In the scattered disk which is the dynamical class of \orten{} 
the typical geometric albedos are between 4\% and 9\%.   


We have recalculated the best fit NEATM models for \orten{} itself by correcting the measured Herschel/PACS flux for the contribution of a satellite with extremely low albedo and beaming parameter. 

In this case the satellite would have \geomalb\,=\,0.02, $\eta$\,=\,0.6, and a correspondig diameter of $\sim$450\,km, resulting in flux densities of 0.99, 1.37 and 1.24\,mJy in the Herschel/PACS 70, 100 and 160\,$\mu$m bands. After correcting for this contribution, the best fit models for \orten{} itself prefer high beaming parameter values of $\eta$\,$\approx$\,2.5, with D$_{\rm eff}$\,$\approx$\,$\sim$1500\,km . However, these high $\eta$ values are very unlikely given the slow rotation of \orten{}. Therefore we also calculated the best fit size of the primary using a fixed beaming parameter value of $\eta$\,=\,1.8, too, the best fit $\eta$ obtained in \citet{Pal2016} (dashed line in Fig.~\ref{fig:thermal}). This provides D$_{\rm eff}$\,=\,1360\,km and a corresponding geometric albedo of \geomalb\,=\,0.11. This size is still larger than the previous estimate for \orten{} by \citet{SS2012} and also that of Haumea \citep[][1240$^{+68.7}_{-58}$\,km]{Fornasier2013}, but smaller than that of Makemake \citep[][1430--1502\,km]{Ortiz2012}.  
We emphasise again that this is an extreme situation any realistic surface assumed for the satellite (\geomalb{}\,$\geq$\,0.04) leaves the \citet{Pal2016} size estimate (D\,$\approx$\,1535\,km) unchanged.

\section{The importance of the satellite of \orten}

Multiple systems are very useful tools for unraveling the main physical properties of trans-Neptunian objects \citep[see e.g.][]{Noll2008}, When diameter measurements are avalialbe, these are the only cases when a reliable estimate of the average density can be obtained. Densities provide information on the internal structure and formation processes 
\citep[][]{Brown2013,Vilenius2014,Grundy2015,Barr2016}.

In a recent paper \citet{Parker2016} reported on a possible discovery of a moon around the dwarf planet Makemake. However, the satellite was identified at a single epoch only. 
Existence of a moon orbiting \orten{} would mean that all known Kuiper belt objects larger than $\sim$1000\,km host satellites, including the four recognized outer dwarf planets: Pluto, Eris, Makemake, Haumea, plus Orcus and Quaoar (the sample discussed in \citet{Barr2016}),
and now \orten. 


While the densities in the additional cases (Makemake and \orten) are not yet known, we can estimate the mass ratios, q, assuming some realistic albedos and near-equal densities. For Makemake the 7\fm0 magnitude difference \citep{Parker2016} results in q\,=\,2$\cdot$10$^{-5}-$5$\cdot$10$^{-4}$, assuming equal or darker albedos for the satellite than that of the primary. For \orten{} equal albedos give q\,=\,0.004, low albedos for the satellite result in q\,$\approx$\,0.01. With these mass ratios all large bodies in our list have q\,$<$0.1 and most systems have q\,$\approx$\,0.01.

Binaries smaller than 1000\,km tend to have nearly equal brightness values, and therefore likely have q\,$>$\,0.1 \citep[see e.g.][for a review]{Noll2008}. 
Near-equal binaries are natural outcome of dynamical capture models \citep[e.g.][]{Astakhov2005} while collisional simulations \citep{Durda2004,Canup2005} can explain the low mass ratios of the satellites of the largest bodies. 
The fact that now \emph{all} Kuiper belt objects with diameters larger than $\sim$1000\,km have satellites underlines the importance of such collisions and may give constraints on the physical conditions in the still dynamically cold disk in the young Solar system. 

With the determination of \orten's satellite's orbit by future observations we will also be able to put constraints on the level of possible tidal dissipation and estimate whether the satellite alone could have slowed down the rotation of \orten{} to the observed $\sim$45\,h value. The bulk density of the \orten{} system would also be of  
significant interest, especially in comparison with that of Makemake, an object of very similar size (D\,$\approx$\,1430km), but with much higher albedo (0.4, vs. 0.09 for \orten) and covered in volatile CH$_4$ ice \citep{Brown2015,Lorenzi2015}. 

\acknowledgements

Data presented in this paper were obtained from the Mikulski Archive for Space Telescopes (MAST). STScI is operated by the Association of Universities for Research in Astronomy, Inc., under NASA contract NAS5-26555. Support for MAST for non-HST data is provided by the NASA Office of Space Science via grant NNX09AF08G and by other grants and contracts. The research leading to these results has received funding from the European Union’s Horizon 2020 Research and Innovation Programme, under Grant Agreement no 687378; from the GINOP-2.3.2-15-2016-00003 grant of the National Research, Development and Innovation Office (Hungary); and from the LP2012-31 grant of the Hungarian Academy of Sciences. Funding from Spanish grant AYA-2014-56637-C2-1-P is acknowledged, as
is the Proyecto de Excelencia de la Junta de Andalucía, J.\,A.\,2012-FQM1776.

\vspace{5mm}
\facilities{HST(STIS)}

\end{document}